\def\Bigskip{\bigskip\bigskip}
\magnification=1200

\centerline{\sl Abstracts of Seminars given at the Workshop on}
\centerline{\bf Mathematical Problems of Quantum Gravity}
\centerline{\sl held at the Erwin Schr\"odinger Institute, Vienna} 
\Bigskip
\Bigskip
\centerline{Peter Aichelburg${}^1$ and Abhay Ashtekar${}^2$}
\centerline{\it Organizers}

\centerline{${}^1$ Institute of Theoretical Physics}
\centerline{University of Vienna, Boltzamanngasse 5, A-1090 Vienna}
\medskip
\centerline{${}^2$ Center for Gravitational Physics and Geometry}
\centerline{Physics Department, Penn State, University Park, PA 16802}

\vskip 7cm

This pre-print contains the abstracts of seminars (including key
references) presented at the ESI workshop on mathematical problems in
quantum gravity held during July and August of 1996. Contributors
include A. Ashtekar, J. Baez, F. Barbero, A. Barvinsky, F. Embacher,
R. Gambini, D. Giulini, J. Halliwell, T. Jacobson, R. Loll, D. Marolf,
K. Meissner, R. Myers, J. Pullin, M. Reisenberger, C. Rovelli,
T. Strobl and T.  Thiemann. While these contributions cover most of
the talks given during the workshop, there were also a few additional
speakers whose contributions were not received in time.

\vfill\break

\centerline{\bf Contributors}
\Bigskip

Ashtekar, Abhay:  	  Ashtekar@phys.psu.edu 

Baez, John     :  	  Baez@.math.ucr.edu

Fernando Barbero:  	  Barbero@LAEFF.ESA.Es

A.O. Barvinsky:	   	  grg@ibrae.msk.edu

Franz Embacher:           fe@pap.univie.ac.at                

Rodolfo Gambini:   	  Rgambini@fisica.edu.uy

Domenico Giulini:  	  DGiulini@esi.ac.at

J.J. Halliwell:           J.halliwell@ic.ac.uk

T. Jacobson:       	  Jacobson@umdhep.umd.edu

Renate Loll:              Loll@aei-potsdam.mpg.de

Don Marolf:        	  Marolf@suhep.phy.syr.edu

Krzysztof A. Meissner:    Meissner@mail.cern.ch

Robert Myers:             RCM@hep.Physics.McGill.CA

Jorge Pullin:		  Pullin@phys.psu.edu

Michael P. Reisenberger:  Mreis@medb.physics.utoronto.ca

Carlo Rovelli:		  Rovelli@vms.cis.pitt.edu

Thomas Strobl:		  Tstrobl@pluto.physik.rrwth-aachen.de

Thomas Thiemann:	  Thiemann@abel.math.harvard.edu

\vfill\break

\centerline{\bf Quantum theory of geometry}
\centerline{\sl Abhay Ashtekar}
\Bigskip

This was primarily a review talk, based largely on joint work with
Jerzy Lewandowski.

Over the last three years, a new functional calculus has been
developed on the quantum configuration space of general relativity
without any reference to a background geometrical structure in
space-time (such as a metric). The purpose of this talk was to
indicate how this machinery can be applied to systematically construct
a quantum theory of geometry. The kinematical Hilbert space of quantum
gravity was presented. States represent polymer-like, 1-dimensional
excitations of geometry. Regulated Operators corresponding to areas of
2-surfaces were introduced on the kinematical Hilbert space of quantum
gravity and shown to be self-adjoint. Their full spectrum was
presented. It is purely discrete and contains some physically
interesting information. First, the ``area gap'', i.e., the value of
the smallest non-zero excitation, contains information about the
global topology of the surface. Second, in the large eigenvalue limit,
the eigenvalues become closer and closer to each other such that
$|a_{n+1} -a_n| \le [(l_P/2\sqrt{a_n}) + O(l_P^2/a_n)] l_P^2$, where
$a_n$ and $a_{n+1}$ are the consecutive eigenvalues and $l_P$ the
Planck length. This shows why the continuum limit is such an excellent
approximation. It also has a more interesting implication for the
Hawking effect. Because we do not have an equal level spacing, the
types of potential problems pointed out by Bekenstein and Mukhanov do
not arise and the semi-classical approximation used by Hawking in his
calculation of the black-body spectrum turns out to be excellent.

These and numerous other results are providing more and more intuition
for the nature of quantum geometry. This framework is to quantum
gravity what familiar differential geometry is to classical general
relativity. Like differential geometry, it has no dynamical content;
specific field equations are not involved. Just as the formulae for
lengths, areas and volumes of differential geometry are valid in all
theories of gravity (involving a space-time metric), the formulae,
identities and theorems involving our quantum states and operators
would hold in all dynamical theories of gravity (which include n-beins
which are canonically conjugate to connections).
\Bigskip
A. Ashtekar and J. Lewandowski, gr-qc/9602046; Class. \& Quantum
Grav. (in press).
\Bigskip
\Bigskip

\centerline{\bf Unforeseen non-commutativity between geometric operators}
\centerline{\sl Abhay Ashtekar}
\Bigskip

This was a report on some joint work with Alejandro Corichi, Jerzy
Lewandowski and Jose Antonio Zapata.

One of the implications of the work reported in the first talk is that
area operators associated with different operators do not always
commute.  This is at first surprising because the classical formula
for areas involves only triads, without any reference to connections
and from the basic Poisson bracket relations one expects the triads to
commute among themselves. It turns out, however, that the naive
expectation is incorrect. The reason is that the formula for areas
involves triads which are smeared {\it only} on 2-surfaces rather than
in 3 dimensions and the Poisson brackets between such objects are,
strictly speaking, singular. (Furthermore, in our framework based on
holonomies, triad operators which are smeared in 3 dimensions are {\it
not} likely to be well-defined!)

To analyze this issue in detail, we examine the Poisson algebra
between the following phase space functions: cylindrical functions of
(smooth) connections and triads smeared on 2-surfaces. (Incidentally,
since the triads have density weight one, they are in fact 2-forms and
it is thus geometrically natural to smear them on 2-surfaces.) If we
assume naively that the smeared triads commute, we run into a problem
with the Jacobi identity; the naive Poisson algebra is not a Lie
algebra and is therefore incorrect. One can regulate the naive algebra
carefully to obtain a Lie algebra. Then, one finds that the
(2-dimensionally) smeared triads fail to Poisson commute. The
commutators between the quantum triad operators just mirror this correct
Poisson algebra and this is why the area operators fail to commute.
\Bigskip
A. Ashtekar, A. Corichi, J. Lewandowski and J.A. Zapata, CGPG pre-print.
\Bigskip
\Bigskip

\centerline{\bf Geometry of quantum mechanics}
\centerline{Abhay Ashtekar}
\Bigskip

This talk summarized joint work with Troy Schilling which constitutes
his 1996 Ph.D. thesis at Penn State.  

In the way we normally formulate these theories, classical mechanics
has deep roots in (symplectic) geometry while quantum mechanics is
essentially algebraic. However, one can recast quantum mechanics in a
geometric language which brings out the similarities and differences
between the two theories. The idea is to pass from the Hilbert space
to the space of rays, i.e. to the ``true'' space of states of quantum
mechanics. The space of rays --or the projective Hilbert space, is in
particular, a symplectic manifold, which happens to be equipped with a
further K\"ahler structure. Regarding it as a symplectic manifold, one
can repeat the familiar constructions from classical mechanics. For
example, given any function, one can construct its Hamiltonian vector
field.  If one uses the expectation value of the Hamiltonian operator
as the function, it turns out that the resulting ``classical''
symplectic evolution is {\it precisely} the (projection of the)
Schr\"odinger evolution on the Hilbert space. Roughly, properties of
quantum mechanics which it ``shares'' with classical mechanics use
only the symplectic structure on the projective Hilbert space. The
``genuinely'' quantum properties such as uncertainties and probabilities
refer to the K\"ahler metric. Thus, purely in mathematical physics
terms, one can regard quantum mechanics as a special case of classical
mechanics, one in which the phase space happens to have a K\"ahler
structure (which then enables one to do more.) This geometrical
formulation of quantum mechanics sheds considerable light on the
second quantization procedure and on semi-classical states and
dynamics.

After the work was completed, we found that many of our results were
discovered independently by a number of authors, most notably L.
Hughstone and by R. Cirelli, A. Mani\'a and L. Pizzochero.
\Bigskip
A. Ashtekar and T. Schilling; in: The Proceedings of the First
Canadian-Mexican-American Physical Societies' Conference, edited by A.
Zapeda (American Institute of Physics, NY 1995.)

T. Schilling; {\sl Geometry of Quantum Mechanics}, Ph.D. Thesis, Penn
State, 1996.  
\Bigskip 
\Bigskip 

\centerline{\bf Probing quantum gravity through} 
\centerline{\bf exactly soluble midi-superspaces}
\centerline{\sl Abhay Ashtekar}
\Bigskip

This talk summarized joint work with Monica Pierri which constituted
part of her Ph.D. thesis.

The idea was to consider midi-superspaces which are simple enough to
be exactly soluble both classically and quantum mechanically and use
the solution to probe various nagging issues of quantum gravity such
as the issue of time and the nature of the vacuum. The specific
example presented was the midi-superspace of Einstein-Rosen (i.e.,
cylindrical) gravitational waves. This model was analyzed by Karel
Kuch\v{a}r already in the early seventies and by Michel Allen in the
mid-eighties. However certain issues concerning boundary conditions,
surface terms and functional analytic subtleties could not be
discussed then. Using results on asymptotics and techniques of
regularization that have been developed since then, their discussion
can be completed to construct a complete quantum theory. We used this
solution to construct a regulated, quantum space-{\it time} metric
operator and address issues such as ``light cone fluctuations''.  That
this is possible within the canonical framework is noteworthy since
concerns are often expressed that the canonical quantization procedure
may not be able to handle such ``space-{\it time}'' issues.  The model
has a well-defined, non-trivial Hamiltonian operator and a {\it
stable} vacuum state (the eigenstate of the Hamiltonian with zero
eigenvalue.)  On general states, one can write the Schr\"odinger
equation.  However, the mathematical parameter in this equation has
the physical interpretation of time only on semi-classical
states. Finally, this solution can also be used to probe a key issue
in our non-perturbative quantum gravity program: existence of
operators corresponding to traces of holonomies around closed
loops. These operators do exist in spite of the fact that they involve
smearing of the connection only in one dimension.

More recently, this model has been used to show the existence of
certain unforeseen quantum gravity effects which can be large even
when the space-time curvature is small.
\Bigskip
A. Ashtekar and M. Pierri, gr-qc/9606085, J. Math. Phys., {\bf 37},
6250-70, (1996).

A. Ashtekar, gr-qc/9610008, Phys. Rev. Lett. {\bf 77}, 4864-67 (1996).

\magnification = 1200
\def\Bigskip{\bigskip\bigskip\goodbreak}
\def\P{{\cal P}}
\def\Q{{\cal Q}}
\def\H{{\cal H}}
\def\S{{\cal S}}

\Bigskip
\Bigskip
\centerline{\bf Topological Quantum Field Theory}
\smallskip
\centerline{\it John Baez}
\Bigskip
The simplest sort of topological quantum field
theory is $BF$ theory, where the Lagrangian is of
the form ${\rm tr}(BF)$, with $F$ being the curvature of
a connection and $B$ being a Lie-algebra valued
$(n-2)$-form in $n$ dimensions.  When $n$ is 3 or 4
one can also add a "cosmological constant term"
of form ${\rm tr}(BBB)$ or ${\rm tr}(BB)$, respectively.  
In this talk, I summarized what is known about $BF$ theory 
in dimensions 2, 3, and 4, as well as the equivalent
state sum models.  In particular, I described how
state sum models of $BF$-like theories in 2 dimensions
arise from certain monoids, while in 3 dimensions 
they arise from certain monoidal categories and in
4 dimensions from certain monoidal 2-categories (most
notably the category of representations of a quantum
group, which may be seen as a monoidal 2-category with
one object).  I also sketched how 4-dimensional $BF$
theory underlies Chern-Simons theory in 3 dimensions.
\Bigskip
References: 
\medskip
John Baez and James Dolan, Higher-dimensional algebra and 
topological quantum field theory, Jour.\ Math.\ Phys.. {\bf 36} 
(1995), 6073-6105.
\medskip
John Baez, Four-dimensional $BF$ theory as a topological 
quantum field theory, to appear in Lett.\ Math.\ Phys.,
preprint available as q-alg/9507006.
\vfill\break

\Bigskip
\Bigskip
\centerline {\bf The Entropy of 2-Part Systems}
\smallskip
\centerline {\it John Baez}
\Bigskip
I sketched the mathematical relationships between
three constructions which might at first glance
seem unrelated: Everett's relative state formalism,
the Gelfand-Naimark-Segal construction, and the 
construction of nontrivial spaces of states on `half'
of a 3-manifold from the single state of $BF$ theory
with cosmological constant on the whole 3-manifold ---
the so-called `Chern-Simons state'.  In all these
constructions a single state gives rise to a Hilbert
space of states.  The last one gives a way of understanding
the mathematical content of Smolin's argument for the
Bekenstein bound.
\Bigskip
References:
\medskip
John Baez, Quantum gravity and the algebra of tangles, 
Jour.\ Class.\ Quantum Grav.\ {\bf 10} (1993), 673-694.
\medskip
Lee Smolin, Linking topological quantum field theory and 
non-perturbative quantum gravity, Jour.\ Math.\ Phys.\ 
{\bf 36} (1995), 6417-6455.
\Bigskip
\Bigskip
\centerline {\bf From Euclidean to Lorentzian Gravity: The Real Way}
\smallskip
\centerline {\it Fernando Barbero}
\Bigskip
The complex character of the Ashtekar variables has been one of the
major issues to be understood in order to succeed in the quantization
of general relativity by using the loop variables approach. The
possibility of describing Lorentzian GR with a real Ashtekar connection
can be realized by restricting oneself only to real canonical
transformations in the transit from the SO(3)-ADM phase space to the
SO(3)-Yang Mills phase space of the Ashtekar formalism [1]. The
resulting Hamiltonian constraint is more complicated that the usual
one but it is still written in terms of Ashtekar variables and, hence,
loop variables can still be used to quantize it. From a Lagrangian
point of view it is interesting to see if one can use local internal
symmetries, instead of non-local ones, to write the action for
Lorentzian general relativity. This can be achieved [2] by writing the
Einstein-Hilbert action for a two parameter family of metrics whose
signature can be adjusted at will by a suitable choice of these
parameters.
\Bigskip

References:

\medskip
J. Fernando Barbero G. {\sl
Phys.Rev.D51:5507-5510, (1995)}.
\medskip

J. Fernando Barbero G.
{\sl Phys.Rev.D54:1492-1499, (1996)}.

\Bigskip
\Bigskip
\centerline {\bf Semiclassical methods in the theory of constrained
dynamics}
\smallskip
\centerline {\it A.O.Barvinsky}

\Bigskip
Operator realization of quantum constraints, the lowest-order structure
functions and physical observables is found in the one-loop
(linear in $\hbar$) approximation of the
Dirac quantization for the general theory subject to first-class constraints.
The general semiclassical solution of the quantum Dirac constraints is found.
Semiclassical unitary equivalence
of the Dirac and reduced phase-space quantization methods is established in
terms of the conserved physical inner product in the space of physical states.
The conservation of this inner product and its independence of the choice
of gauge conditions is based on the Stokes theorem for a special closed
form integrated over the physical subspace of the configuration space
of the theory (superspace). Geometrical covariance properties of the
quantum Dirac constraints with respect to contact canonical transformations
and transformations of the basis of constraints is studied. Applications
of these general methods of quantum constrained dynamics are considered in
quantum cosmology of the early inflationary Universe.

\Bigskip
References:
\medskip
A.O. Barvinsky, The general semiclassical solution of the
Wheeler-DeWitt equations and the issue of unitarity in quantum
cosmology, Phys. Lett. {\bf B241} (1990) 201.
\medskip
A.O. Barvinsky and V. Krykhtin, Dirac and BFV quantization
methods in the 1-loop approximation: closure  of the quantum constraint
algebra and the conserved inner product, Class. Quantum Grav. {\bf 10}
(1993) 1957.
\medskip
A.O.Barvinsky, Operator ordering in theories subject to
constraints of the gravitational type, Class. Quantum Grav. {\bf 10}
(1993) 1985.
\medskip
A.O.Barvinsky, Unitarity approach to quantum cosmology,
Phys.Reports {\bf 230} (1993) 237.
\medskip
A.O.Barvinsky, Geometry of the Dirac quantization of
constrained systems, preprint ESI (1996).

\Bigskip
\Bigskip
\vfill\break
\centerline {\bf Quantum Origin of the Early Universe and the
Energy Scale of Inflation}
\smallskip
\centerline{\it A.O.Barvinsky}

\Bigskip
Quantum origin of the early inflationary Universe from the no-boundary
and tunnelling quantum states is considered in the one-loop
approximation of quantum cosmology. A universal effective action
algorithm for the distribution function of chaotic inflationary
cosmologies is derived for both of these states.  The energy scale of
inflation is calculated by finding a sharp probability peak in this
distribution function for a tunnelling model driven by the inflaton
field with large negative constant $\xi$ of nonminimal
interaction. The sub-Planckian parameters of this peak (the mean value
of the corresponding Hubble constant {\bf H}$\simeq
10^{-5}m_P$, its quantum width $\Delta${\bf H}/{\bf
H}$\simeq 10^{-5}$ and the number of inflationary e-foldings
{\bf N}$\simeq 60$) are found to be in good correspondence with
the observational status of inflation theory, provided the coupling
constants of the theory are constrained by a condition which is likely
to be enforced by the (quasi) supersymmetric nature of the
sub-Planckian particle physics model.

\Bigskip

References:
\medskip
A.O.Barvinsky and A.Yu.Kamenshchik, One-loop
quantum cosmology: the normalizability of the Hartle-Hawking
wave function and the probability of inflation, Class. Quantum
Grav.{\bf 7} (1990) L181.
\medskip
A.O.Barvinsky, Unitarity approach to quantum cosmology,
Phys.Reports {\bf 230} (1993) 237.
\medskip
A.O.Barvinsky and A.Yu.Kamenshchik, Tunnelling geometries:
analyticity, unitarity and instantons in quantum cosmology,
Phys.Rev. {\bf D50} (1994) 5093.
\medskip
A.O.Barvinsky, Reduction methods for functional
determinants in quantum gravity and cosmology,
Phys.Rev. {\bf D50} (1994) 5115.
\medskip
A.O.Barvinsky and A.Yu.Kamenshchik, Quantum scale of
inflation and particle phy\-sics of the early Universe,
Phys.Lett. {\bf B332} (1994) 270.

\Bigskip
\Bigskip
\vfill\break

\centerline{\bf Mode decomposition and unitarity in quantum cosmology}
\centerline{\sl Franz Embacher}
\Bigskip

\medskip

It is common folklore that the space of wave functions of quantum
cosmology may not be decomposed into positive and negative frequency
modes when the background structure (DeWitt metric and potential) does
not admit symmetries.  However, there are still perspectives for
defining a generalized notion of preferred mode decomposition,
starting from the space of solutions of the wave equation and the
indefinite Klein-Gordon scalar product $Q(\psi_1,\psi_2)= -{{i}\over
{2}} \int_\Sigma d\Sigma^\alpha\, (\psi_1^* {\nabla}_\alpha \psi_2 -
\psi_2^* \nabla_\alpha \psi_1)$.  In the case of a positive potential
$U$ we outline a strategy in doing so.

The technical tool for analyzing the wave equation is the selection of
a solution $S$ of the Hamilton-Jacobi equation, which generates a
congruence of classcial trajectories, and a weight function $D$. The
pair $(S,D)$ is called WKB-branch.  Within any WKB-branch, an operator
$H$ ---satisfying a particular differential equation--- may be chosen
such that any solution of $i\partial_t\chi=H\chi$ (with $\partial_t$
the derivative along the trajectories) gives rise to a solution of the
original wave equation $(-\nabla^2+U)\psi=0$.  The wave functions
constructed in this way define the space $\H^+$, generalizing the
notion of positive frequency with respect to a WKB-branch.

The crucial step in the general strategy is to perform
a {\it natural} choice of the operator $H$ with respect to any
WKB-branch, such that for any two infinitesimally close branches
the respective decompositions {\it coincide}.
It is achieved by iteratively solving the differental equation for $H$, 
thus ending up with a formal expression 
(whose existence at least in simple cases can be inferred explicitly). 
Since arbitrarily high derivatives of the ingredients of the 
model (the DeWitt metric and the potential) appear, the 
proper mathematical existence of the preferred decomposition 
is likely to be related to the global structure of the model 
(and possibly to analyticity issues). 
Unitarity (i.e. a Schr{\"o}dinger type evolution
equation for wave functions) shows up only in terms of 
WKB-branches, much like components of a tensor show up only
in a coordinate system. Details of this approach are presented in the two
references cited below, along with speculations on a possible
relation to the refined algebraic quantization program. 

This work was supported by the Austrian Academy of Sciences in the
framework of the ''Austrian Programme for Advanced Research and
Technology.''
\Bigskip

F. Embacher, Decomposition and unitarity in quantum cosmology ,
{\it preprint}\hfil\break UWThPh-1996-64, gr-qc/9611006; 

F. Embacher, Mode decomposition and unitarity in quantum cosmology,
Talk given at the {\it Second Meeting on Constrained Dynamics and
Quantum gravity}, Santa Margherita Ligure, September 17--21, 1996, to
appear in the Proceedings.  

\Bigskip
\Bigskip

\centerline {\bf Gauge Invariance in the Extended Loop Representation}
\Bigskip
\centerline {\it Rodolfo Gambini}
\Bigskip
\Bigskip
The Gauss constraint in the extended loop representation is studied.  It 
is shown that there is a sector of the state space that is gauge 
invariant.  We determine necessary and sufficient conditions for states 
belonging to this sector.  This conditions are satisfied by the extended 
Vassiliev invariants.
\Bigskip

References:
\medskip
C. Di Bartolo, R. Gambini, J. Griego, J. Pullin J Math. Phys. 36 
6510, 1995.
\medskip
C. Di Bartolo, "The Gauss Constraint in the Extended Loop 
Representation" prepring gr-qc 9607014, 1996.

\Bigskip
\Bigskip

\centerline{\bf Views on Super-selection Rules in QT and QFT}
\smallskip

\centerline{\it Domenico Giulini}
\Bigskip

\def\P{{\cal P}}
\def\Q{{\cal Q}}
\def\H{{\cal H}}

Many derivations of super-selection rules are purely formal in nature
and hence do not make sufficiently clear the actual physical input
that leads to them. In the context of quantum mechanics we critically
review standard derivations of the super-selection rules for
univalence and overall mass (so-called Bargmann super-selection rule.)
The strong dependence on the required symmetry group is emphasized
[1]. In particular, it is pointed out that in order for mass to define
a super-selection rule it should be considered as a dynamical
variable. We present a minimal extension of the dynamics of $n$ point
particles interacting via some Galilei invariant potential in which
mass it also treated as dynamical variable. Here the classical
symmetry group turns out to be given by an $R$-extension of the
Galilei group. In contrast to the Galilei group its extension does
have an action on the space of states including non-trivial
superpositions of mass eigenstates. Hence no super-selection rules
appear in this model [2].  Finally, we discuss the super-selection
rule of electric charge (see chap. 6 of [1]).  We emphasize the
necessity of a consistent variational principle including charged
configurations in its domain of differentiability.  We present such a
principle and show the unavoidable existence of additional degrees of
freedom (``surface variables'') which measure the overall multipole
moments of the charge distribution.  Here super-selection rules result
only if the surface variables are not in the algebra of observables,
as it would be the case if one restricts to (quasi-) local
observables. However, it is tempting to think of such a restriction as
being conditioned dynamically and hence in principle only of
approximate validity [3].

\Bigskip

References:
\medskip
D.~Giulini, E.~Joos, C.~Kiefer, J.~Kupsch, O.~Stamatescu,
H.D.~Zeh: ``Decoherence and the Appearance of a Classical
World in Quantum Theory'', Springer Berlin (1996).
\medskip
D. Giulini: ``On Galilei invariance in Quantum Mechanics
and the Bargmann Super-selection Rule'', {\it Ann. Phys. (NY)}
{\bf 249},  (1996)
\medskip
D. Giulini, C. Kiefer, H.D. Zeh: ``Decoherence, Symmetries
and Super-selection Rules, {\it Phys. Lett. A} {\bf 119}, 291-298
(1995).
\Bigskip
\Bigskip

\def\P{{\cal P}}
\def\Q{{\cal Q}}
\def\H{{\cal H}}

\centerline{\bf Diffeomorphism Invariant Subspaces in}

\centerline{\bf Witten's 2+1-Dimensional Quantum Gravity on 
$T^2\times R$}
\smallskip

\centerline{\it Domenico Giulini}


\Bigskip
We address the r\^ ole of large diffeomorphisms in Witten's
formulation of gravity as an ISO(2,1) gauge theory on a space-time
$T^2\times R$. In a ``space-like sector'' the classical phase space is
$\P=T^*(\Q )$ with $\Q=R^2-\{0\}$ as configuration space. Using the
vertical polarization the quantum state space becomes $\H=L^2(\Q,
\mu_L)$, $\mu_L$ being the standard Lebesgue measure. By large
diffeomorphisms we mean the orientation preserving mapping class group
of $T^2$, given by $SL(2,Z)$. It acts by its defining representation
on $\Q$ and by its canonical lift on $\P$. The action on the former is
`wild' in the sense that the isomorphicity class of stabilizer
subgroups is nowhere locally constant. The configuration space
quotient is hence nowhere locally a manifold. In contrast, the lifted
action on $\P$ is free on the open and dense subset where the two
vectors of coordinates and momenta are not perpendicular.  The action
of $SL(2,Z)$ on $\H$ is given by composition in the argument of the
function representing the element in $\H$.  We show that the action is
fully reducible in terms of a direct integral over $R$ of vector
spaces isomorphic to $L^2(S^2,d\varphi)$, which carry
$SL(2,Z)$-irreducible sub-representations of certain irreducible
representations of $SL(2,R)$ from the continuous series [1].  Hence
for each measurable set $\Delta\subset R$ one finds a closed subspace
$\H_{\Delta}\subset \H$ which is invariant under large
diffeomorphisms. Now, if $SL(2,Z)$ is considered as gauge group, the
observables should be contained in its commutant, implying in
particular that no vector in $\H$ defines a pure state for this
algebra. $\H$ is not the physical Hilbert space nor does it contain
it. We conclude that the reduction of large diffeomorphisms should a
priori not be regarded as a problem simpler than the reduction of
diffeomorphisms generated by the constraints (identity component).
\Bigskip

References:
\medskip
D.~Giulini, J.~Louko: ``Diffeomorphism Invariant Subspaces in
Witten's 2+1 Quantum Gravity on $R\times T^2$'', {\it Class. Quant.
Grav.} {\bf 12}, 2735-2745 (1995).

\Bigskip
\Bigskip

\centerline{\bf Mapping-Class Groups of General 3-Manifolds}

\smallskip
\centerline{\it Domenico Giulini}

\def\S{{\cal S}}
\Bigskip
Let $\Sigma$ be a closed orientable 3-manifold, $\infty\in\Sigma$ a
distinguished point, and $D_F(\Sigma)$ the space of diffeomorphisms
that fix the frames at $\infty$. We are interested in the mapping
class groups $\S(\Sigma):=D_F(\Sigma)/D_F^0(\Sigma)$ for general
$\Sigma$, where $D_F^0(\Sigma)$ denotes the identity component of
$D_F(\Sigma)$ [1]. As is well known, $\Sigma$ is diffeomorphic to a
connected sum of finitely many (say $n$) and uniquely determined prime
3-manifolds $P_i$. [NB: $P$ is prime $\Leftrightarrow$ $\pi_2(P)=0$ or
$P=S^2\times S^1$ (the `handle').] Using this, we think of $\Sigma$ as
a configuration of $n$ elementary `objects' attached to a common base
along mutually disjoint connecting spheres [2].  An obvious subgroup
of $\S(\Sigma)$ is generated by the semi-direct product of
permutations of diffeomorphic objects with their internal symmetry
groups $\S(P_i)$. This subgroup is in fact also a factor iff $P_i\not
= S^2\times S^1$ [3].  It is explained how a full presentation of
$\S(\Sigma)$ may be constructed using the Fuks-Rabinovich presentation
for the automorphism group of free products [3].  For this one
considers the natural map $h:\ \S(\Sigma)\rightarrow
\hbox{Aut}(\pi_1(\Sigma,\infty))$.  If all $P_i$ satisfy that
homotopic diffeomorphisms are also isotopic (no prime violating this
is known), the kernel, $\hbox{ker}(h)$, is known to be of the form
$Z_2^m$, $m\leq n$, generated by certain rotations parallel to
imbedded 2-spheres.  The image, $\hbox{Im}(h)$, can be explicitly
presented. In many cases it exhausts all of
$\hbox{Aut}(\pi_1(\Sigma))$. Finally, one specifies the correct action
of $\hbox{Im}(h)$ on $\hbox{ker}(h)$, which results in a semi-direct
product of these two groups. All this can be nicely exemplified
explicitly for the connected sum of $n$ handles or $n$ $PR^3$'s
[2][3].

\Bigskip

References:
\medskip
D.~Giulini: ``On the Configuration Space Topology in General
Relativity'', {\it Helv. Phys. Acta} {\bf 68}, 86-111 (1995).
\medskip
D.~Giulini: ``3-Manifolds for Relativists'', {\it Int.Jour. Theor.
Phys.} {\bf 33}, 913-930 (1994).
\medskip
D. Giulini: ``The Group of Large Diffeomorphisms in General
Relativity. {\it Banach Center Publications}, in press.
\Bigskip
\vfill\break

\centerline {\bf On the Probability of Entering a Spacetime Region}
\centerline{\bf in Non-Relativistic Quantum Mechanics}
\smallskip
\centerline {\it J.J.Halliwell and E.Zafiris}
\Bigskip
\Bigskip
What is the probability of a particle entering a given region of space
at any time between $t_1$ and $t_2$? Standard quantum theory assigns
probabilities to alternatives at a fixed moment of time and is not
immediately suited to questions of this type. We use the decoherent
histories approach to quantum theory to compute the probability of a
non-relativistic particle entering a spacetime region.  Aside from
being of general formal interest, this question is relevant to the
problems of arrival times and tunneling times. It may also be relevant
to relativistic systems and in particular, to quantum gravity, where a
variable playing the role of time may not exist. For a system
consisting of a single non-relativistic particle, histories
coarse--grained according to whether or not they pass through
spacetime regions are generally not decoherent, except for very
special initial states, and thus probabilities cannot be
assigned. Decoherence may, however, be achieved by coupling the
particle to an environment consisting of a set of harmonic oscillators
in a thermal bath.  Probabilities for spacetime coarse grainings are
thus calculated by considering restricted density operator propagators
of the quantum Brownian motion model, and we find approximate methods
for calculating these.  Another method of achieving decoherence, which
we explore, is to consider a system consisting of a large number $N$
of identical, non-interacting, free particles, and consider histories
in which an imprecisely defined proportion of the particles cross the
spacetime region. We find that there is decoherence, essentially due
to statistics for large $N$. We thus obtain general expressions for
the probabilities for a variety of spacetime problems for a particle
starting in an arbitrary initial state.
\Bigskip
\Bigskip

\centerline {\bf Issues in Black Hole Thermodynamics}
\smallskip 
\centerline {\it T. Jacobson}
\Bigskip
This talk provided an overview of black hole
thermodynamics and discussed recent progress and open
questions. The issues discussed  included: the generalized
second law, entanglement entropy,  the ``holographic
hypothesis", and the statistical meaning of black hole
entropy. In particular, the relation between matter field
contributions to the entropy and  the renormalization of
Newton's constant, the nature of the ``bare" entropy, and
Carlip's  counting of black hole states in 2+1-dimensional
quantum gravity were discussed.
\Bigskip
References:
\medskip
Susskind, L., ``The world as a hologram", 
Jour. Math. Phys. 36 (1995) 6377.
\medskip
Kabat, D., Shenker, S.H. and Strassler, M.J., ``Black hole entropy
in the O(N) model", Phys. Rev. D 52 (1995) 7027.
\medskip
Carlip, S., ``The statistical mechanics of the three-dimensional 
euclidean black hole", gr-qc/9606043.
\medskip

\Bigskip
\Bigskip
\centerline {\bf Origin of the Outgoing Black Hole Modes}
\smallskip
\centerline {\it T. Jacobson}

\Bigskip
The origin of the outgoing black hole  modes is
puzzling if no transplanckian reservoir at the horizon is
availavble. In this talk I explained this puzzle and
discussed models with high frequency dispersion, motivated
by condensed matter analogies, which can resolve the
puzzle.  These models arose originally from Unruh's sonic
analog of  a black hole, which is an inhomogeneous  fluid
flow with a ``sonic horizon" where the flow speed exceeds
the speed of sound. I explained how high frequency
dispersion in the  wave equation satisfied by the sound
field (or its analog) leads to a process of ``mode
conversion", whereby ingoing short wavelength modes are
converted into long wavelength  outgoing ones. Results of a
calculation of  the Hawking spectrum in one such model were
described.

\Bigskip
References:
\medskip
Unruh, W.G.,  ``Sonic analog of black holes and
the effects of high frequencies on black hole evaporation",
Phys. Rev. D 51 (1995) 2827.
\medskip
Corley, S. and Jacobson, T.,
``Hawking spectrum and high frequency dispersion", Phys.
Rev. D 54 (1996) 1568-1586.
\medskip  
Jacobson, T., ``On the
origin of the outgoing black hole modes", Phys. Rev. D 53
(1996) 7082-7088.
\Bigskip
\Bigskip
\vfill\break

\centerline {\bf 1+1 Sector of 3+1 Gravity}
\smallskip
\centerline {\it T. Jacobson}

\Bigskip
Ashtekar's formulation of general relativity
admits an extension to  degenerate metrics, and it appears
that such metrics may play an important role in the
quantization of the theory. Recently Matschull showed how
this degenerate extension of GR can be described in a
fully  spacetime covariant manner, and he showed that the
degenerate ``geometries" allowed by the Ashtekar extension
always possess a local ``causal cone", though the cone is
collapsed in one or more dimensions.  In this talk, the
rank-1 sector of the {\it classical} theory was discussed,
motivated by the  degeneracy of the triad along the Wilson
lines in quantum loop states.  I showed that the classical
lines behave like (1+1)-dimensional spacetimes with a pair
of massless Dirac fields propagating along them
(``connection waves"). Matschull's causal structure is
precisely the light cone for these waves. Further, if the
lines form a  congruence of closed loops, the holonomy must
be the same on all the loops. Results for inclusion of
matter and supergravity were also obtained in this work.

\Bigskip
References:
\medskip
Matschull, H.J., 
``Causal structure and diffeomorphisms in Ashtekar's gravity",
Class. Quantum Grav. 13 (1996) 765-782.
\medskip
Jacobson, T., ``1+1 sector of 3+1 gravity", 
Class. Quantum Grav. 13 (1996) L111-L116.
\Bigskip
\Bigskip
\centerline {\bf Some News of Lattice Gravity}
\smallskip
\centerline {\it R. Loll}

\Bigskip
At the Vienna workshop, I reported about some new results which I
obtained in the quantization of Hamiltonian lattice gravity.  Because
of the close structural resemblance of the calculations, these are
also of relevance to the continuum loop quantization.

I am working with (a discretized version of) real connection variables
and their canonically conjugate momenta $(A,E)$ on a cubic
$N^3$-lattice.  The natural scalar product at the kinematical level is
therefore identical with that of the usual lattice gauge theory with a
compact gauge group ($SO(3)$ or $SU(2)$), and the basic link operators
(the link holonomy $\hat U(l)$ and link momenta $\hat p_i(l)$) are
self-adjoint. The Hamiltonian constraint is non-polynomial, but this
can be handled along the lines described in [1].

Like in the continuum, one may define self-adjoint operators measuring
volumes and areas, and the gauge-invariant states diagonalizing these
operators are simple linear combinations of so-called spin network
states. The one-dimensional Wilson loop states underlying this
construction are obtained by simply multiplying together the
one-dimensional basic link holonomies (that form part of the single,
fixed lattice), whereas in the continuum they are somewhat less
natural composite objects depending on the connections $A$ and on
arbitrary {\it embedded} loops in the three-dimensional spatial
manifold $\Sigma$. Related to this is the fact that in the continuum
theory -- unlike on the lattice -- there still exist natural ${\rm
Diff}\,\Sigma$-actions.

The first interesting property I found is the following: suppose one
wanted to lattice-quantize the classical phase space function
corresponding to the ``area of a surface perpendicular to the
3-direction", $\int d^2x\,\sqrt{E^{3i}E^3_i}$. To reproduce the
spectrum found by Ashtekar and Lewandowski in the continuum, one
substitutes the continuum momenta $E^b(x)$ by symmetrized lattice
momenta ${1 \over 2} (p^+(n,b)+p^-(n,b))$, where $n$ is now a lattice
vertex and the symmetrization is taken over the link momenta in
positive and negative $b$-direction, both based at (i.e. transforming
non-trivially at) $n$. Doing this, I realized that this operator is
not diagonal in terms of certain volume eigenstates I had constructed
previously on the lattice, that is, volume and area operators in
general do not commute! (The commutator may still vanish on certain
``simple" loop states.) I was at first greatly worried by this result,
since the corresponding classical phase space functions do of course
commute. Moreover, the calculations I did can be directly translated
to the continuum, which means that also there areas and volumes do not
commute. Details of this can be found in my forthcoming paper [2]. On
the lattice there is an easy way around this problem: choose a
different discretization for the term under the square root, namely,
${1\over 2} (p^+(n,b)^2+p^-(n,b)^2))$ instead of ${1\over 4}
(p^+(n,b)+p^-(n,b))^2$.  This differs from the latter by terms of
higher order in the lattice spacing $a$, as $a\rightarrow 0$ in the
continuum limit, and is therefore an equally good operator from the
point of view of the lattice theory. Being a sum of two laplacians, it
commutes strongly with {\it any} other lattice operator.

I came across another curious feature while looking for simultaneous
eigenvectors of the volume and the area operators on the lattice.
Since their operator expressions reduce to sums over vertex
contributions, one can diagonalize them separately at each
intersection. The dual unit cubes around individual vertices may
therefore be regarded as smallest building blocks of geometry.  I
looked at a particular family of states at some fixed vertex $n$,
namely those where the six links meeting at $n$ (the lattice is cubic)
have the same occupation number (or spin) $j=1,2,3,\dots$, and only
differ by how the flux lines are contracted gauge-invariantly at
$n$. One may then extract local length scales by computing
$\sqrt{a_0}$ and $\root 3 \of{v_0}$, with $a_0$ and $v_0$ the
eigenvalues of the area (which for these states are the same in all
three directions) and the volume. One finds that at least for the
first few $j$ the length scale one obtains from the area is larger
than that calculated from the volume, even if one always picks the
state of highest volume from the entire set. For example, for $j=1$,
one finds $\sqrt{a}\sim 0.866$, $\root 3 \of{v_{\rm max}}\sim 0.821$,
and for $j=2$, $\sqrt{a}\sim 1.189$, $\root 3 \of{v_{\rm max}}\sim
1.077$, in suitable units. This behavior seems to persist also for
higher $j$ [3]. It remains to be understood to what extent this is a
general feature of local lattice states. If it were, it would be
difficult to understand how one could construct states representing
flat space, say, from those smallest building blocks.
\vfill\break
\Bigskip
References:
\medskip

R. Loll: ``A Real Alternative to Quantum Gravity in Loop Space",
to appear in {\it Phys. Rev. D}.
\medskip
R. Loll, in preparation.
\medskip
R. Loll, in preparation.
\Bigskip
\Bigskip

\centerline {\bf D-branes and Black Hole entropy}
\smallskip
\centerline {\it Donald Marolf}
\Bigskip
A review was presented of recent calculations of black hole entropy in
string theory, giving the general picture of how calculations of
D-brane bound states are related to black hole entropy.  The original
calculation by Strominger and Vafa (Phys. Lett. B379, 99 (1996);
hep-th/9601029) was outlined and some extensions were mentioned.  A
discussion along these lines and additional references can be found in
the review by Horowitz, gr-qc/9604051.

\Bigskip
\centerline {\bf Duality Symmetries in String and Field Theory}
\smallskip
\centerline {\it Krzysztof A. Meissner}
\Bigskip
The talk given at the ESI Workshop on Quantum Gravity in Vienna
described duality symmetries both in string and in field theories. The
subject is now being intensively studied for at least three
reasons. The first one is that one aspect of duality (strong vs. weak
coupling) gives us hope to probe normally inaccessible region of
strongly coupled quantum field theories by showing their equivalence
with (usually different) weakly coupled theories where the
perturbation expansion can be trusted. This kind of equivalence was
shown up to now in a limited number of theories (like N=2
supersymmetric Yang-Mills theories) but even these few examples are
extremely helpful in understanding strongly interacting quantum field
theories.  The second reason is that they relate seemingly different
string theories (like heterotic and IIB) which is seen upon
compactifying them on two different manifolds and comparing the
resulting effective actions. The number of dualities of that type is
rapidly increasing and it led to speculations that all string theories
are interconnected and in fact they all descend from one theory
(called ``M-theory'' if 11-dimensional or ``F-theory'' if 12
dimensional) but compactified with different boundary conditions.  The
third reason that dualities are intensively studied is that they are
``solution generating'' symmetries i.e. starting with a solution to
the equations of motion of a given theory and acting on it with
elements of the duality group, we get a whole class of new solutions.
The resulting solutions can be very complicated and rather impossible
to get by solving the equations of motion. Such global symmetries have
also a conserved current which is very helpful in classifying the
solutions. One particular example is the so called string cosmology
with gravity coupled to a scalar (dilaton) and antisymmetric tensor
where there is an $O(d,d)$ global symmetry where $d$ is the number of
space dimensions. This symmetry allows for many generalizations and is
always present in the gravitational sector of ``string-inspired''
effective actions.

\Bigskip
\Bigskip

\centerline {\bf Black Hole Entropy from Strings and D-Branes}
\smallskip

\centerline{\it Robert Myers}
\Bigskip
\Bigskip

Finding a statistical mechanical interpretation of black hole entropy
is an outstanding problem which has eluded physicists for over 20
years.  Recently, progress into this question has been made using new
insights from string theory. This progress is a spin-off from the work
on string dualities, and the realization of the important role of
extended objects beyond just strings. In particular, a class of
extended objects known as D-branes [1] have proven very valuable from a
calculational standpoint.  It was found that different kinds of
D-branes can be combined to produce black holes in a certain strong
coupling limit. On the other hand in weak coupling, these systems are
amenable to statistical mechanical analysis within string theory.
These calculations were first carried out for a class of extremal
black holes in five-dimensions [2], but then rapidly extended to a
variety of other configurations [3-7].  Even though these calculations
still apply to a relatively restricted class of black holes, they
represent a breakthrough in our understanding of black hole entropy,
since for the first time, we have some insight into the underlying
microscopic degrees of freedom for a black hole.
\Bigskip
References:
\medskip

J.~Polchinski, S.~Chaudhuri and C.V.~Johnson,
``Notes on D-branes'', hep-th/9602052.
\medskip
A.~Strominger and C.~Vafa, ``Microscopic Origin of the
Bekenstein-Hawking
Entropy'', {\it Physics Letters} {\bf B379} (1996) 99
[hep-th/9601029].
\medskip
C.G.~Callan and J.M.~Maldacena, ``D-Brane Approach to Black
Hole Quantum Mechanics'', {\it Nuclear Physics} {\bf B472} (1996)
591 [hep-th/9602043];
\medskip
G.T.~Horowitz and A.~Strominger, ``Counting States of Near Extremal
Black Holes'', {\it Physical Review Letters} {\bf 77}
(1996) 2368 [hep-th/9602051].
\medskip
J.C. Breckenridge, R.C. Myers, A.W. Peet and C. Vafa,
``D-Branes and Spinning Black Holes'',  hep-th/9602065;
\medskip
J.C. Breckenridge, D.A. Lowe, R.C. Myers, A.W. Peet,
A. Strominger and C. Vafa, ``Macroscopic and Microscopic
Entropy of Near Extremal Spinning Black Holes'',
{\it Phys. Lett.} {\bf B381} (1996) 423
[hep-th/9603078].
\medskip
C.V. Johnson, R.R. Khuri and R.C. Myers,
``Entropy of 4-D Extremal Black Holes'',
{\it Phys. Lett.} {\bf B378} (1996) 78 [hep-th/9603061];
\medskip
J.M. Maldacena and A. Strominger, ``Statistical Entropy of
Four-Dimensional Extremal Black Holes'',
{\it Phys. Rev. Lett.} {\bf 77} (1996) 428 [hep-th/9603060].
\medskip
G.T. Horowitz, J.M. Maldacena and A. Strominger,
``Nonextremal Black Hole Microstates and U-Duality'',
hep-th/9603109;
\medskip
G.T. Horowitz, D.A. Lowe and J.M. Maldacena,
``Statistical Entropy of Nonextremal Four-Dimensional
Black Holes and U-Duality'',
{\it Phys. Rev. Lett.} {\bf 77} (1996) 430 [hep-th/9603195].
\medskip
G.T. Horowitz and D. Marolf, ``Counting States of Black Strings
with Traveling Waves'', hep-th/9605224;
\medskip
``Counting States of Black Strings
with Traveling Waves 2'', hep-th/9606113.

\Bigskip
\Bigskip
\centerline {\bf Quantum Gravi-dynamics as skein relations in knot space}
\smallskip
\centerline {\it Jorge Pullin}
\Bigskip
In the loop representation, states that are solutions of the
diffeomorphism constraint are knot invariants. Typically, one tries to
find a realization of the Hamiltonian constraint that acts on such
states in order to find quantum states of the gravitational
field. Such an action can never yield an operator in the space of
knots, since the Hamiltonian is a non diffeomorphism invariant
function of a point, and therefore cannot be realized in a space of
diffeomorphism invariant states. We argue, however, that many
proposals for regularized actions of the Hamiltonian can be decomposed
into a non diffeomorphism invariant pre-factor times a topological
operator. The latter can be realized in a space of diffeomorphism
invariant wave functions. We analyze in particular a recently proposed
lattice regularization [1]. In terms of it, the topological operator
can be interpreted as a skein relation between intersecting knots
[2]. This relation determines partially the knot polynomial that is
the general solution of the Einstein equations. The indeterminacy is
related to the fact that the theory does not have a single solution
since it has local degrees of freedom.  We show that certain knot
invariants, which in the continuum [3] and extended [4] loop
representations were found to solve formally the Hamiltonian
constraint satisfy the skein relations found to represent the dynamics
of general relativity, providing additional confirmation that they
could be states of gravity. The calculations are carried out for a
particular type of triple intersections; further studies will be
needed to elucidate in a more general way if the states are actually
compatible with the skein relations for general intersections. The
idea of viewing the constraint as partial skein relations is not
confined to this particular approach and holds promise in the context
of the recently proposed Hamiltonian for real Lorentzian gravity in
terms of spin network states.
\Bigskip
References:
\medskip
R. Gambini, J. Pullin, "The general solution of the quantum
Einstein equations?" preprint gr-qc/9603019
\medskip
H. Fort, R. Gambini, J. Pullin, "Lattice knot theory and quantum
gravity in the loop representation", preprint gr-qc/9608033
\medskip
[3] B. Bruegmann, R. Gambini, J. Pullin, Phys. Rev. Lett 68 431 (1992).
\medskip
[4] C. Di Bartolo, R. Gambini, J. Griego, J. Pullin, Phys. Rev. Lett. 
72 3297 (1994).
\Bigskip
\Bigskip

\centerline {\bf A left-handed simplicial action for euclidean general 
relativity} 
\smallskip
\centerline {\it Michael P. Reisenberger}
\Bigskip

An action for simplicial euclidean general relativity involving only 
left-handed fields is presented. The simplicial theory is shown to converge to
continuum general relativity in the Plebanski formulation as the simplicial 
complex is refined. 
\medskip
An entirely analogous hyper-cubic lattice theory, which approximates
Plebanski's form of general relativity is also presented.

\vfill\break
\Bigskip
\Bigskip
\centerline{\bf A path integral formulation of}
\centerline{\bf  loop quantized simplicial euclidean general relativity} 
\Bigskip
\centerline{\it Michael P. Reisenberger}
\Bigskip

A four dimensional path integral formulation of simplicial euclidean
general relativity (GR) corresponding to canonical GR in Ashtekar's
connection representation is presented.  By integrating out the
spacetime connection the path integral is turned into a sum over
spins, analogous to the Ponzano-Regge model for $2+1$ GR,
corresponding to canonical GR in the spin network representation. In
this latter case the path integral may be interpreted as a) a sum over
world sheets of spin networks, and b) a sum over discrete spacetime
geometries analogous to the discrete spatial geometries found in loop
quantized canonical GR in the continuum. The discreteness of the
4-geometry should persist in a continuum limit of the simplicial model
because it results from the discreteness of the spins of $SU(2)$
representations, not from the discreteness of the simplicial complex
modeling spacetime.  However, the existence of a continuum limit has
not been established.

The path integral model is derived from a new classical simplicial
action which in the classical continuum limit converges to the
Plebanski action for GR.
\Bigskip
References:
\medskip

M.~Reisenberger.  A left-handed simplicial action for euclidean
general relativity.  {\it gr-qc 9609002}, 1996.  (presents the
classical simplicial theory.)

\Bigskip
\Bigskip
\centerline{\bf Black Hole Entropy from Loop Quantum Gravity}
\smallskip
\centerline{\it Carlo Rovelli}
\Bigskip
I have discussed recent ideas on the possibility of deriving the 
Bekenstein-Hawking formula, which states that the Entropy of a
(non rotating) black hole is proportional to its Area, from Loop
Quantum Gravity. 
\Bigskip
References:
\medskip
C Rovelli, ``Loop Quantum Gravity and Black Hole Physics'', 
gr-qc/9608032. Contains an introduction to Loop Quantum Gravity, and a 
detailed discussion.
\medskip
C Rovelli, ``Black Hole Entropy from Loop Quantum Gravity'', 
gr-qc/9603063. Short: only the main idea and computation. 
\medskip
K Krasnov, ``On statistical mechanics of gravitational systems'' 
gr-qc/9605047. A slightly different approach.
\medskip
K Krasnov, ``The Bekenstein bound and non-perturbative quantum gravity'', 
gr-qc/9603025.
\Bigskip
\Bigskip

\centerline{\bf Lessons from 1+1 Gravity}
\smallskip
\centerline{\it T. Strobl}
\Bigskip

Subject of our investigations is the wide class of generalized 2d
dilaton gravity Lagrangians
$$ L[g,\Phi] = \int d^2 x \; \sqrt{|\det g|} \,\left[U(\Phi) \, R +
  V(\Phi) + W(\Phi) \; \partial_\mu \Phi \partial^\mu \Phi \right]
\, \, , $$
where $U,V,W$ are functions of the dilaton $\Phi$.  I briefly reviewed
what is known about these (midi-superspace) models on the classical
level [1]. In particular, for Lorentzian signature of the metric $g$
{\it all}\/ classical, diffeomorphism inequivalent solutions have been
found. For not too specific choices of $U,V$, and $W$ these include
perfectly smooth solutions on {\it any}\/ ('reasonable') non-compact
two-surface as well as various multi black hole configurations.

I then came to discuss the Dirac quantization of (1). Here the
reformulation of this action in terms of so-called Poisson
$\sigma$-models [2], comparable to the formulation of 2+1 gravity as
Chern-Simons theory, is essential. In the context of (1) the target
space of the $\sigma$-model is an $R^3$, equipped with a Poisson
bracket induced by the choice of $U,V$, and $W$.  Correspondingly this
auxiliary $R^3$ foliates (stratifies) into (generically)
two-dimensional symplectic sub-manifolds, characterized by a target
space coordinate $M$. {\it On-shell}\/ the latter may be identified
with the (generalized) mass of the spacetime, a Dirac observable of
the theory. The general framework of Poisson $\sigma$-models allows to
determine the spectrum of $M$ by means of a simple analysis of the
above foliation; e.g., $Spec(M)$ is discrete, {\it iff}\/ the
respective symplectic leaves have non-trivial second homotopy. For
some choices (but not for all) of $U,V,W$ in (1) $Spec(M)$ becomes
discrete for Euclidean signature of the theory (i.e.\ of the metric
$g$), but continuous for Lorentzian signature.  If a similar relation
holds for {\it some}\/ (any) Dirac observable $O$ of 4d Einstein
gravity, the recently proposed generalized Wick transformation
$O_{Lor} = T O_{Eucl} T^{-1}$ cannot hold, at least in a strict sense.

In his treatment of the 2+1 black hole Carlip assigns the black hole
entropy to states of a WZW boundary action located at the (stretched)
horizon. In the 1+1 dimensional context of (1) such a boundary action
will describe purely mechanical degrees of freedom. The following nice
picture evolves [3]: The phase space of these edge degrees of freedom
on a black hole spacetime of mass $M$ coincides precisely with the
respective symplectic leaf in the above mentioned $R^3$; thus the
fictitious point particles obtained above become 'alive' and
physical. Despite this appealing picture, the entropy obtained in this
way does {\it not}\/ seem to agree with the one obtained by other,
quite reliable semiclassical approaches.

\Bigskip

References:
\medskip
T.\ Kl\"osch and T.\ Strobl,  {\it Class.\ Quantum Grav.}
13 (1996) 965-983, 2395-2421, and hep-th/9607226. 
\medskip
P.\ Schaller and T.\ Strobl, {\it Mod.\ Phys.\ Letts.}
{\bf A9} (1994), 3129 as well as proceedings contributions 
 hep-th/9411163 and  hep-th/9507020. 
\medskip
J.\ Gegenberg, G.\ Kunstatter, and T.\ Strobl,
gr-qc/9607055 and in preparation. 

\vfill\break
\Bigskip
\Bigskip

\centerline {\bf Quantum Spin Dynamics}
\smallskip
\centerline {\it Thomas Thiemann}
\Bigskip

An anomaly-free operator corresponding to the Wheeler-DeWitt
constraint of Lorent\-zian, four-dimensional, canonical,
non-perturbative vacuum gravity is constructed in the continuum. This
operator is entirely free of factor ordering singularities and can be
defined in symmetric and non-symmetric form.

We work in the real connection representation and obtain a
well-defined quantum theory. We compute the complete solution to the
Quantum Einstein Equations for the non-symmetric version of the
operator and a physical inner product thereon.

The action of the Wheeler-DeWitt constraint on spin-network states is
by annihilating, creating and rerouting the quanta of angular momentum
associated with the edges of the underlying graph while the ADM-energy
is essentially diagonalized by the spin-network states. We argue that
the spin-network representation is the ``non-linear Fock
representation" of quantum gravity, thus justifying the term ``Quantum
Spin Dynamics (QSD)".
\Bigskip
References:
\medskip
 
T. Thiemann, ``Anomaly-free formulation of non-perturbative, four-dimensional
Lor\-entzian quantum gravity, HUTMP-96/B-350, gr-qc/9606088,
Phys. Lett. B380 (1996) 257-264
\medskip
T. Thiemann, ``Quantum Spin Dynamics (QSD)", Harvard
Preprint HUTMP-96/B-351, gr-qc/9606089
\medskip
T. Thiemann, ``Quantum Spin Dynamics (QSD) II", Harvard
Preprint HUTMP-96/B-352, gr-qc/9606090
\Bigskip
\Bigskip
\vfill\break
\centerline {\bf A Length Operator for Canonical Quantum Gravity}
\smallskip
\centerline {\it Thomas Thiemann}
\Bigskip
We construct an operator that measures the length of a curve in
four-dimensional Lorentzian vacuum quantum gravity.

We work in a representation in which a $SU(2)$ connection is diagonal
and it is therefore surprising that the operator obtained after
regularization is densely defined, does not suffer from factor
ordering singularities and does not require any renormalization.

We show that the length operator admits self-adjoint extensions and
compute part of its spectrum which like its companions, the volume and
area operators already constructed in the literature, is purely
discrete and roughly is quantized in units of the Planck length.

The length operator contains full and direct information about all the
components of the metric tensor which faciliates the construction of
so-called weave states which approximate a given classical 3-geometry.
\Bigskip
References :
\medskip
T. Thiemann, ``A length operator in canonical quantum 
gravity", Harvard Preprint HUTMP-96/B-354, gr-qc/960692
\Bigskip

\end
\bye